# Recovering the lost steerability of quantum states within non-Markovian environments by utilizing quantum partially collapsing measurements

Wen-Yang Sun [1], Dong Wang [1, 2], Zhi-Yong Ding [1, 3] and Liu Ye [1, *]

[1] *School of Physics & Material Science, Anhui University, Hefei 230601, People's Republic of China*
[2] *CAS Key Laboratory of Quantum Information, University of Science and Technology of China, Hefei 230026, People's Republic of China*
[3] *Research center of Quantum Information Technology, School of Physics & Electronic Engineering, Fuyang Normal University, Fuyang 236037, People's Republic of China*

**Abstract:** In this Letter, we mainly investigate the dynamic behavior of quantum steering and how to effectively recover the lost steerability of quantum states within non-Markovian environments. We consider two different cases (one-subsystem or all-subsystem interacts with the dissipative environments), and obtain that the dynamical interaction between system initialized by a Werner state and the non-Markovian environments can induce the quasi-periodic quantum entanglement (concurrence) resurgence, however, quantum steering cannot retrieve in such a condition. And we can obtain that the resurgent quantum entanglement cannot be utilized to achieve quantum steering. Subsequently, we put forward a feasible physical scheme for recovering the steerability of quantum states within the non-Markovian noises by prior weak measurement on each subsystem before the interaction with dissipative environments followed by post weak measurement reversal. It is shown that the steerability of quantum states and the fidelity can be effectively restored. Furthermore, the results show that the larger the weak measurement strength is, the better the effectiveness of the scheme is. Consequently, our investigations might be beneficial to recover the lost steerability of quantum states within the non-Markovian regimes.

# 1. Introduction

The existence of quantum correlations in the mixed quantum systems is widely recognized as one of the most fundamental features of the quantum information theory, which can distinguish the quantum realm from the classical one [1-3]. Conceptually, entanglement [4-6], as a particular kind of quantum correlation, can be quantified by the Wootters' concurrence [7]. Besides, entanglement, as an important information resource, is able to fulfill a variety of realistic tasks of quantum information processing (QIP) [8-11]. On the other hand, quantum steering can be used as an important resource in QIP as well. The phenomenon of quantum steering was firstly introduced by Schrödinger in 1935 to interpret Einstein-Podolsky-Rosen paradox [12, 13]. Then, some theoretical and experimental works regarding quantum steering have been achieved [14-21], and Wiseman *et al.* [22, 23] formulated steering in an operational way in conformity for one quantum information task.

---

* **Corresponding author: yeliu@ahu.edu.cn**

Recently, quantum steering was given an operational explanation as the distribution of entanglement by an untrusted party [23, 24]. And it has a one-way property [16], that is, one bipartite quantum state may be steerable from *A* to *B*, however *B* cannot steer *A*. Generally speaking, quantum steering is considered as an intermediate type of quantum correlations between entanglement [5] and Bell nonlocality [25] in the region of modern quantum information theory. In addition, quantum steering can be effectively detected by violating quantum steering inequalities [26, 27]. Derived for both discrete and continuous variable systems [28-31], such some inequalities of quantum steering can be obtained by utilizing entropy uncertainty relation (EUR) [30, 31]. Some promising criteria for quantum steering have been observed to probe steering on various aspects [32-37].

However, these previous explorations were nearly limited to investigate quantum steering in an isolated quantum system. The systems in essence are disclosed and unavoidably interact with their surrounding noises under the realistic regimes, giving rise to decoherence that exponentially damages the entanglement [38, 39]. Additionally, it has been demonstrated that the quantum correlation in an open system is characterized by some especial phenomena, such as entanglement sudden death and revival of entanglement [40-42]. Mazzola *et al.* [43] have found that the revival of the entanglement (concurrence) after a finite time period of the entire absence can be achieved when the system is interacted with a canonical non-Markovian regime [44, 45]. Naturally there is an interesting question that whether the state during revival for quantum entanglement can steer under the regime of non-Markovian environments. To our best knowledge, there have been a few authors to exploit the steerability of quantum states in different noisy channels [46-48]. Herein, we will focus on resolving the issue under the influence of non-Markovian regimes.

To be explicit, we will consider two different cases (one-subsystem or all-subsystem interacts with the non-Markovian environments), and mainly investigate the dynamic behaviors of quantum entanglement and steering. It turns out that the dynamical interaction between quantum states and noise can induce quasi-periodic quantum entanglement revival, however, quantum steering cannot obtain revival. And the resurgent quantum entanglement cannot be used to realize steering. Furthermore, we explore a scenario of recovering the steerability of quantum states within non-Markovian environments. It shows that our scenario can successfully recover the destroyed entanglement and the lost steerability of Werner state by employing quantum partially collapsing measurements. What's more, our scenario can remarkably recover fidelity as well.

The Letter is organized as follows. In Sect. 2, the bipartite *X*-state for quantum steering is briefly reviewed. In Sect. 3, we explore the dynamic behavior of steering in non-Markovian environments and how to recover the steerability of the quantum states. In Sect. 4, we end up our paper with a concise conclusion.

## 2. Quantum steering of bipartite *X*-state

It is well known that the *X*-shaped quantum states of two-qubit can be expressed as

$$\rho^X = \begin{pmatrix} \rho_{11} & 0 & 0 & \rho_{14} \\ 0 & \rho_{22} & \rho_{23} & 0 \\ 0 & \rho_{23} & \rho_{33} & 0 \\ \rho_{14} & 0 & 0 & \rho_{44} \end{pmatrix}, \quad (1)$$

where $\rho_{ij}(i, j = 1, 2, 3, 4)$ are all real parameters. By using appropriate local unitary transformations, one can rewrite the state $\rho^X$ in Bloch decomposition

$$\rho^X = \frac{1}{4}\begin{pmatrix} 1+c_3+s+r & 0 & 0 & c_1-c_2 \\ 0 & 1-c_3+r-s & c_1+c_2 & 0 \\ 0 & c_1+c_2 & 1-c_3-r+s & 0 \\ c_1-c_2 & 0 & 0 & 1+c_3-r-s \end{pmatrix}, \quad (2)$$

with

$$\begin{aligned} & c_1 = 2(\rho_{23}+\rho_{14}), \ c_2 = 2(\rho_{23}-\rho_{14}), \ c_3 = \rho_{11}-\rho_{22}-\rho_{33}+\rho_{44}, \\ & r = \rho_{11}+\rho_{22}-\rho_{33}-\rho_{44}, \ s = \rho_{11}-\rho_{22}+\rho_{33}-\rho_{44}. \end{aligned} \quad (3)$$

Here, the EUR steering inequality is introduced. If the quantum state can violate the EUR

steering inequality, the quantum state is steerable. According to Refs. [28-31], by applying the positivity of the continuous relative entropy between any couple of probability distributions, Walborn *et al.* [28] argued that it is always the case for continuous observables (COs) in states allowing local hidden states (LHS) models $h(x^B|x^A) \geq \int d\lambda \, \rho(\lambda) h_q(x^B|\lambda)$, where $h_q(x^B|\lambda)$ is the continuous Shannon entropy caused by probability density. It is shown (as Walborn *et al.* did) that any state allowing a LHS model in position and momentum must satisfy [30]

$$h(x^B|x^A) + h(k^B|k^A) \geq \log(\pi e). \tag{4}$$

Note that here and throughout the paper the base of all logarithms is deemed to be 2. Subsequently, one notes that the same arguments used to develop LHS constraints for COs can be employed to formulate LHS constraints for discrete observables (DOs) as well [30]. Because the positivity of the relative entropy is a fact for both continuous and discrete variables [49], one can derive the corresponding LHS constraint for DOs in the same way: $H(R^B|R^A) \geq \sum_\lambda P(\lambda) H_q(R^B|\lambda)$, where $H_q(R^B|\lambda)$ is the discrete Shannon entropy of $P_q(R^B|\lambda)$. Therewith, one immediately obtains a new entropic steering inequality [30]

$$H(R^B|R^A) + H(S^B|S^A) \geq \log(\Omega^B), \tag{5}$$

where $\Omega^B$ is the value $\Omega \equiv \min_{i,j}\left(1/|\langle R_i|S_j\rangle|^2\right)$, $\{|R_i\rangle\}$ and $\{|S_j\rangle\}$ are the eigenbases of observables $\hat{R}^B$ and $\hat{S}^B$ in the same *N*-dimensional Hilbert space [50], respectively. We must realize that for any EUR, even some relating more than two observables, there is a corresponding steering inequality [30]. In 2-D quantum systems, employing the Pauli *X*, *Y* and *Z* measurements bases on each side, one can obtain the quantum steering if the condition [30, 31]

$$H\left(\sigma_x^B|\sigma_x^A\right) + H\left(\sigma_y^B|\sigma_y^A\right) + H\left(\sigma_z^B|\sigma_z^A\right) \geq 2 \tag{6}$$

is violated, where $H(B|A) = H(\rho_{AB}) - H(\rho_A)$ is the conditional von Neumann entropy. For simplicity, by using Eqs. (2) and (6), the steering inequality of the bipartite *X*-state can be given by

$$\sum_{i=1,2}\left[(1+c_i)\log(1+c_i) + (1-c_i)\log(1-c_i)\right] - (1+r)\log(1+r) - (1-r)\log(1-r)$$
$$+ \frac{1}{2}\left[(1+c_3+r+s)\log(1+c_3+r+s) + (1+c_3-r-s)\log(1+c_3-r-s)\right. \tag{7}$$
$$\left. + (1-c_3-r+s)\log(1-c_3-r+s) + (1-c_3+r-s)\log(1-c_3+r-s)\right] \leq 2.$$

Similar to the Bell-CHSH inequality violation, if the steering inequality (7) is violated, the quantum steering appears. Therewith, choosing a convenient normalization, we can define quantum steering to quantify how much the general bipartite *X*-state is steerable by Pauli measurements

$$S := \max\left\{0,\ \frac{SI - 2}{SI_{\max} - 2}\right\}, \qquad (8)$$

with

$$\begin{aligned}SI := &\sum_{i=1,2}\left[(1+c_i)\log(1+c_i) + (1-c_i)\log(1-c_i)\right] - (1+r)\log(1+r) - (1-r)\log(1-r) \\ &+ \frac{1}{2}\big[(1+c_3+r+s)\log(1+c_3+r+s) + (1+c_3-r-s)\log(1+c_3-r-s) \\ &+ (1-c_3-r+s)\log(1-c_3-r+s) + (1-c_3+r-s)\log(1-c_3+r-s)\big],\end{aligned} \qquad (9)$$

and $0 \leq SI \leq SI_{\max} = 6$ (when the state is a maximal entangled pure state, $SI_{\max} = 6$), one guarantees that $S \in [0, 1]$.

## 3. Dynamic behavior of quantum steering of bipartite states within non-Markovian environments

### 3.1 Non-Markovian regimes

Considering a system of two qubits, each interacts only with its local noise independently at zero temperature whose dynamics is derived by the Jaynes-Cummings model. The dynamics of the total system can be obtained directly from the evolution of the individual pairs. The single "qubit + reservoir" Hamiltonian [42] is given by

$$H = \omega_0 \sigma_+ \sigma_- + \sum_i \omega_i b_i^\dagger b_i + (\sigma_+ B + \sigma_- B^\dagger), \qquad (10)$$

where $B = \sum_i g_i b_i$ with $g_i$ being the coupling constant, $\omega_0$ denotes the transition frequency of the two-level qubit and $\sigma_\pm$ are the raising and lowering operators. Besides, the index $i$ labels the field modes of the noise with frequency $\omega_i$ and $b_i^\dagger (b_i)$ is the modes' creation (annihilation) operator. In addition, the Hamiltonian in Eq. (10) may describe a variety of systems [42], such as one qubit formed by an exciton in a potential that represents one quantum well. Herein, the system as a qubit interacting with a non-Markovian reservoir will be considered, and the qubit is composed of a two-level atom. In Ref. [51], the local non-Markovian noise at zero temperature can be quantized as an electromagnetic field in a high-$Q$ cavity with the following effective spectral density

$$J(\omega) = \frac{\gamma_0 \lambda^2}{2\pi[(\omega_0 - \omega)^2 + \lambda^2]}, \quad (11)$$

where the parameters $\gamma_0$ and $\lambda$ define the decay rate of the excited state of the qubit and the spectral width of the noise, respectively. Particularly, their relative magnitudes typically determine a non-Markovian ($\lambda < 2\gamma_0$) and a Markovian ($\lambda > 2\gamma_0$) regime, respectively [52]. We will focus on the non-Markovian regime in terms of $\lambda = 0.1$ and $\gamma_0 = 1$ in the remainder of our work. From Ref. [53], the general notation $\{|0\rangle, |1\rangle\}$ is used as the computational basis of the qubit. One discovers that the action of decoherence noise over single qubit $S$ can be represented by the following quantum map [54]

$$\begin{aligned}|0\rangle_S |0\rangle_E &\to |0\rangle_S |0\rangle_E, \\ |1\rangle_S |0\rangle_E &\to \sqrt{G_t} |1\rangle_S |0\rangle_E + \sqrt{1 - G_t} |0\rangle_S |1\rangle_E.\end{aligned} \quad (12)$$

The dynamical map for the single qubit $S$ evolved within the non-Markovian regime can be also described by the reduced density matrix [54-56]

$$\rho^S(t) = \begin{pmatrix} \rho_{00}^S(0) + \rho_{11}^S(0)(1 - G_t) & \rho_{01}^S(0)\sqrt{G_t} \\ \rho_{10}^S(0)\sqrt{G_t} & \rho_{11}^S(0) G_t \end{pmatrix}, \quad (13)$$

where $G_t$ is an oscillation term describing the fact that the decay of the qubit' excited state is induced by the interaction between the single qubit system and the noise. For the effective spectral density $J(\omega)$ in Eq. (11), within the non-Markovian noise, $G_t$ can be given by [53-56]

$$G_t = e^{-\lambda t} \left[ \cos\left(\frac{dt}{2}\right) + \frac{\lambda}{d} \sin\left(\frac{dt}{2}\right) \right]^2, \quad (14)$$

where $d = \sqrt{2\gamma_0 \lambda - \lambda^2}$. It is shown that in the non-Markovian regime $G_t$ starts to oscillate, which will lead to a non-monotonic behavior of the system state.

### 3.2 How to recover the lost steerability of a bipartite quantum state

Suppose that there is a bipartite system composed of two qubits. The two-qubit state is initially prepared in a Werner state

$$\rho_{AB} = p|\varphi^-\rangle\langle\varphi^-| + \frac{1-p}{4}I, \tag{15}$$

where $|\varphi^-\rangle = (|01\rangle - |10\rangle)/\sqrt{2}$ is a maximally entangled state and $0 \leq p \leq 1$. Note that Werner state is an entangled state in the case of $1/3 < p \leq 1$, while is a disentangled state for $0 \leq p \leq 1/3$.

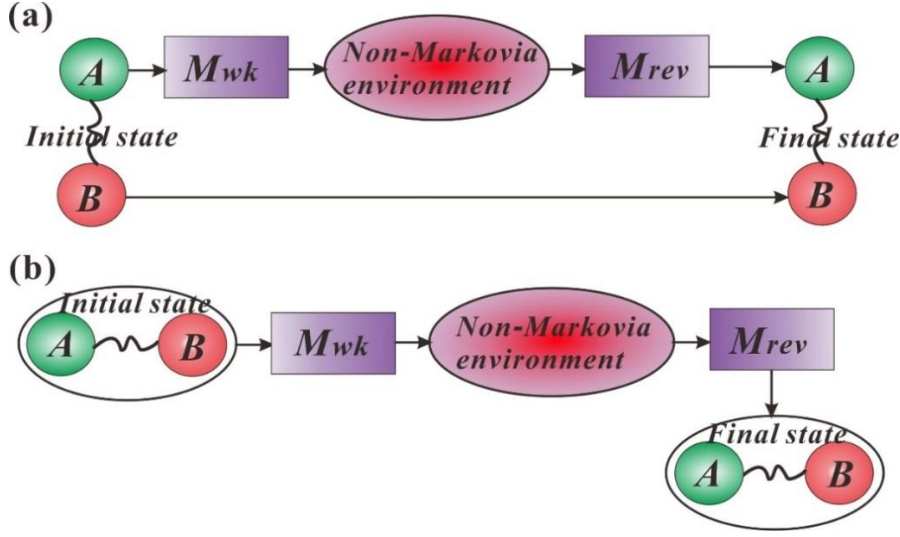

**Fig. 1 (a)** The subsystem *A* interacts with a non-Markovian environment, and the process for recovering quantum steering employing weak measurement and measurement reversal. We call case 1. **(b)** The system is coupled with the non-Markovian environment, and the process for recovering the steerability of quantum state using weak measurement and measurement reversal. We call the other case.

Herein, we consider two canonical cases, the case 1: the one *B* is isolated while the other *A* locally interacts with a dissipative environment. The other case, the two qubits couple with a common local dissipative environment. In addition, for recovering the steerability of the states, the weak measurement (WM) and the weak measurement reversal (WMR) are employed before and after the dissipative environment, respectively. For clarity, the physical model sketch of the total system is depicted in Fig. 1, where the dissipative environment is turned to the non-Markovian regime. It is widely known that WM and WMR are quantum partially collapsing measurements [57-59]. The WM and WMR operations for a single qubit can be expressed as [48, 60]

$$M_{wk} = |0\rangle\langle 0| + \sqrt{1-m}|1\rangle\langle 1|, \tag{16}$$

$$M_{rev} = \sqrt{1-m_r}|0\rangle\langle 0| + |1\rangle\langle 1|, \tag{17}$$

respectively, where $m$ is the strength of WM, and $m_r$ is the WMR strength. Additionally, as is well known, the degree of entanglement for two-qubit system can be quantified conveniently by concurrence. We here chose the concurrence as an entanglement witness. The concurrence is defined as [7] $C = \max\{0, \sqrt{\lambda_1} - \sqrt{\lambda_2} - \sqrt{\lambda_3} - \sqrt{\lambda_4}\}$, and $\lambda_1 \geq \lambda_2 \geq \lambda_3 \geq \lambda_4 \geq 0$, where $\lambda_i (i = 1, 2, 3, 4)$ are the eigenvalues of the matrix $R = \rho(\sigma_y \otimes \sigma_y)\rho^*(\sigma_y \otimes \sigma_y)$. If the density matrix is X-structure, there is a reduced form for concurrence shown as [61]

$$C = 2\max\{0, |\rho_{14}| - \sqrt{\rho_{22}\rho_{33}}, |\rho_{23}| - \sqrt{\rho_{11}\rho_{44}}\}, \tag{18}$$

where $\rho_{ij}$ are the elements of the matrix $\rho^X$ for Eq. (1).

### 3.2.1 The subsystem A locally interacts with a non-Markovian environment

The case 1 is shown as Fig. 1 (a): a qubit B is isolated while the other A locally interacts with the non-Markovian noise for recovering quantum steering by employing WM and WMR. Through these operations, the non-zero elements of the final density matrix $\rho_{(a)m_r}$ are

$$\rho_{(a)m_r 11} = \frac{(m_r - 1)[2 + G_t(m-1)(1+p) - m(1+p)]}{2(m - 2 + (2 + G_t(m-1) - m)m_r)},$$

$$\rho_{(a)m_r 22} = \frac{(1 - m_r)[m - 2 + G_t(m-1)(p-1) - mp]}{2(m - 2 + (2 + G_t(m-1) - m)m_r)},$$

$$\rho_{(a)m_r 33} = \frac{G_t(m-1)(1+p)}{2(m - 2 + (2 + G_t(m-1) - m)m_r)}, \tag{19}$$

$$\rho_{(a)m_r 44} = \frac{G_t(1-m)(p-1)}{2(m - 2 + (2 + G_t(m-1) - m)m_r)},$$

$$\rho_{(a)m_r 23} = \frac{p\sqrt{G_t}\sqrt{1-m}\sqrt{1-m_r}}{m - 2 + (2 + G_t(m-1) - m)m_r} = \rho_{(a)m_r 32}.$$

Employing the Eq. (18), the expression of concurrence can be obtained

$$C_{(a)} = \max\left[0, \frac{\sqrt{\Theta} - 2p\sqrt{G_t}\sqrt{1-m}\sqrt{1-m_r}}{m - 2 + (2 + G_t(-1+m) - m)m_r}\right], \tag{20}$$

where $\Theta = G_t(1-m_r)(m-1)(p-1)[2 + (1+p)(G_t(m-1) - m)]$. Then, by using Eqs. (3) and (19), the corresponding each parameter expression of the bipartite damped states (BDS) in Bloch decomposition can be obtained. It is straightforward to insert each parameter of the BDS $\rho_{(a)m_r}$ into Eqs. (8) and (9), we can obtain analytical expression of quantum steering.

Now, we have at hand two control parameters $m$ and $m_r$, which we can manage to manipulate the qubit's correlation for a certain purpose at any time during the evolution. If we would like to get the best effect for quantum correlation recovery, one needs an optimal WMR strength. According to Refs. [60, 62, 63], by solving equations, we can obtain the optimal WMR strength which gives the maximum amount of quantum correlation

$$m_{(a)or} = \frac{2 - 2G_t - m + 2G_t m}{2 - G_t - m + G_t m}. \tag{21}$$

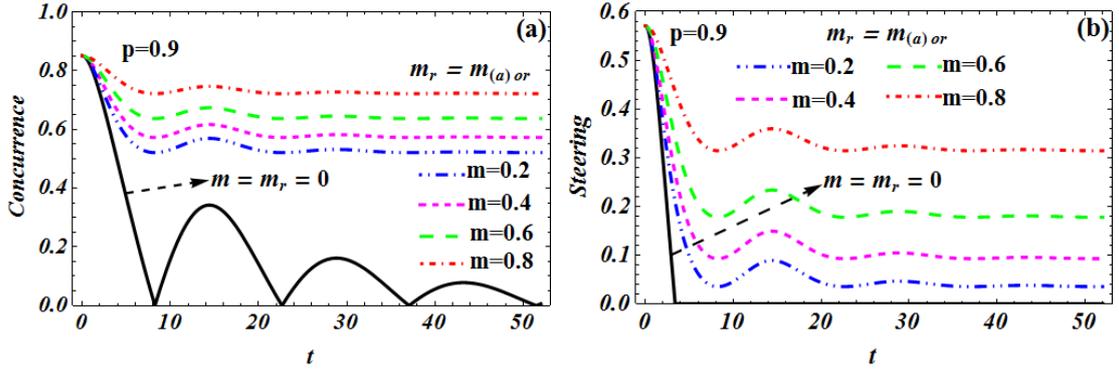

**Fig. 2** (Color online) The quantum correlations as a function of the time parameter $t$ for the different values of the WM $m$ with $m_r = m_{(a)or}$. Here, **(a)** shown the entanglement (concurrence), and **(b)** shown the quantum steering.

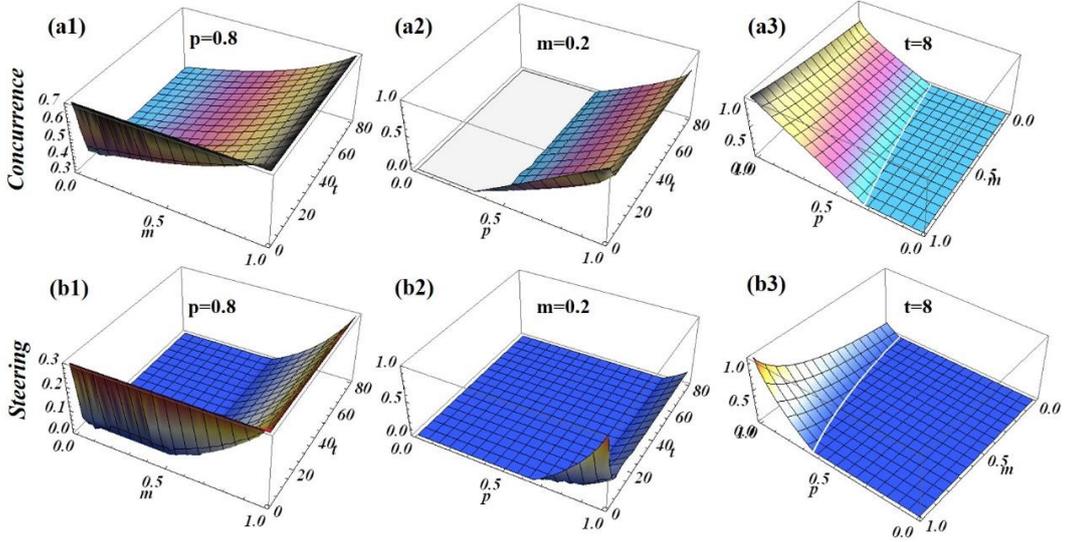

**Fig. 3** (Color online) The quantum correlations versus $m$ and $t$ within the non-Markovian regime for $p = 0.8$; The quantum correlations versus $p$ and $t$ for $m = 0.2$; The quantum correlations versus $p$ and $m$ for $t = 8$. **(a1)**, **(a2)** and **(a3)** shown concurrence, **(b1)**, **(b2)** and **(b3)** shown steering.

As shown in Fig. 2, if we consider that the WM and WMR do not work. The dynamical interaction between quantum state and noise can induce the quasi-periodic quantum entanglement (concurrence) revival, while quantum steering cannot obtain revival. And the redivious quantum entanglement cannot be employed to realize steering. Then, when we choose the optimal WMR strength for Eq. (21), the entanglement can be enhanced with increasing WM strength, and the enhanced entanglement can also be employed to realize steering when the WM strength is big enough.

In order to better explore the quantum correlations how depend on the parameters ($m$, $p$, $t$) for $m_r = m_{(a)or}$, we draw the Fig. 3. In the figure, one can see that quantum entanglement and steering increase with the increase of the WM strength, and generally decrease with the increase of the time parameter $t$. We also obtain that entanglement and steering increase with the increase of the state parameter $p$. Here, one can find that the WM can better recover quantum entanglement and steering, and maybe better recover the initial state by employing the optimal WMR strength. Meanwhile, when quantum steering and entanglement just appear, the state parameter $p$ has a critical value, the critical value is $p \approx 0.65$ and $p = 1/3$, respectively. That is, when the state parameter is approximately larger than 0.65, the damped state possible has an ability that it can be used to realize steering. If the state parameter is larger than $1/3$ but less than 0.65, the damped state may be unsteerable and only entangled, when the time parameter $t$ is smaller. And the state parameter $p$ is less than $1/3$, the damped state is disentanglement. These results may be useful to analyze the distribution of quantum correlations (quantum entanglement and steering) for the Werner states in the non-Markovian regime.

### 3.2.2 The system is coupled with the non-Markovian environments

Next, as shown in Fig. 1 (b), the other case is shown that the system couples with one common local dissipative environment, and the process for recovering the steerability of quantum state by utilizing WM and WMR operations. Through the above these operations, the non-zero elements of the density matrix $\rho_{(b)m_r}$ are

$$\rho_{(b)m_r 11} = \frac{(m_r - 1)^2 \left[ m(4 + m(p-1)) - 4 + G_t^2(p-1) + G_t(4 + m(m - 4 - mp)) \right]}{Q},$$

$$\rho_{(b)m_r 22} = \frac{G_t(G_t - 2 + m - pG_t + mp)(1 - m_r)}{Q},$$

$$\rho_{(b)m_r 33} = \frac{G_t(m-1)(m - 2 + G_t(m-1)(p-1) - mp)(m_r - 1)}{Q},$$

$$\rho_{(b)m_r 44} = \frac{G_t^2 (1-m)^2 (p-1)}{Q}, \qquad (22)$$

$$\rho_{(b)m_r 23} = \frac{2G_t(m-1)p(m_r - 1)}{Q} = \rho_{(b)m_r 32}.$$

with

$$\begin{aligned}
Q &= 2m\big[2 + (G_t(2 + G_t - G_t p - 2m_r) + 2(-2 + m_r))m_r\big] \\
&\quad - 4 - m^2(-1 + p)\big[(1 - G_t)(2 + G_t - m_r)m_r - 1\big] \\
&\quad + m_r\big[8 - 4m_r + G_t(-4 + (4 + G_t(-1 + p))m_r)\big].
\end{aligned} \qquad (23)$$

Employing the Eq. (18), we can obtain the expression of concurrence

$$C_{(b)} = \max\left[0, \frac{2\sqrt{\Upsilon} - 4G_t(-1+m)p(m_r - 1)}{Q}\right], \qquad (24)$$

where

$$\Upsilon = G_t^2(1-m)^2(p-1)\big[m(4 + m(p-1)) - 4 + G_t^2(p-1) + G_t(4 + m(m - 4 - mp))\big](m_r - 1)^2.$$

By using Eqs. (3) and (22), one can obtain the corresponding each parameter expression of the BDS in Bloch decomposition. Then, it is straightforward to insert each parameter of the BDS $\rho_{(b)m_r}$ in Bloch decomposition into Eqs. (8) and (9), one can obtain the analytical expression of quantum steering.

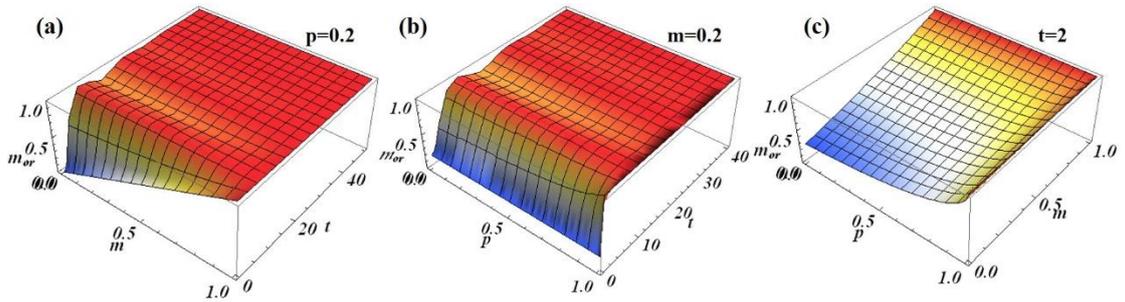

**Fig. 4** (Color online) **(a)** The optimal WMR strength $m_{or}$ versus $m$ and $t$ for $p = 0.2$; **(b)** The optimal WMR strength versus $p$ and $t$ for $m = 0.2$; **(c)** The optimal WMR strength versus $p$ and $m$ for $t = 2$.

In the same manner as Sect. 3.2.1, the corresponding optimal WMR strength which gives the maximum amount of quantum correlation approximately can be obtained

$$m_{(b)or} = 1 - \sqrt{\frac{G_t^2(1-m)^2(p-1)}{m(4+m(p-1))-4+G_t^2(p-1)+G_t(4+m(m-4-mp))}}. \quad (25)$$

The optimal strength of the WMR as a function of different parameters ($m$, $p$, $t$) is shown in Fig. 4. It is obvious that our method does not violate common sense because the value of the optimal WMR strength always ranges from 0 to 1.

Assumed that the WMR strength is optimal, we primarily display how quantum entanglement and steering vary with $t$ and $m$ in Fig. 5. It is shown that both entanglement and steering of the given state can be remarkably recovered, we also discover the larger the WM strength, the better the entanglement and steering enhancement. As shown in Fig. 6, one can obtain that the steerability of the damped state increases with the increase of the WM strength and $p$, but decreases with the increase of the time parameter $t$. In addition, one can see that the damped state can be employed to realize steering as well, if and only if the state parameter is approximately larger than 0.65. These results effectively testified that our scheme can remarkably recover the steerability of Werner state within the non-Markovian regime. If we do not use the WM and WMR operations, quantum entanglement has a quasi-periodic revival, while quantum steering does not have revival. And the resurgent quantum entanglement cannot be employed to realize steering. Through these results and the analysis of the Sect. 3.2.1, we can conclude that not every entangled state can be applied to realize steering, even if the entanglement value reaches a fixed boundary value. In this sense, we can conjecture that the boundaries of their entanglement with steering are very different for different entangled states.

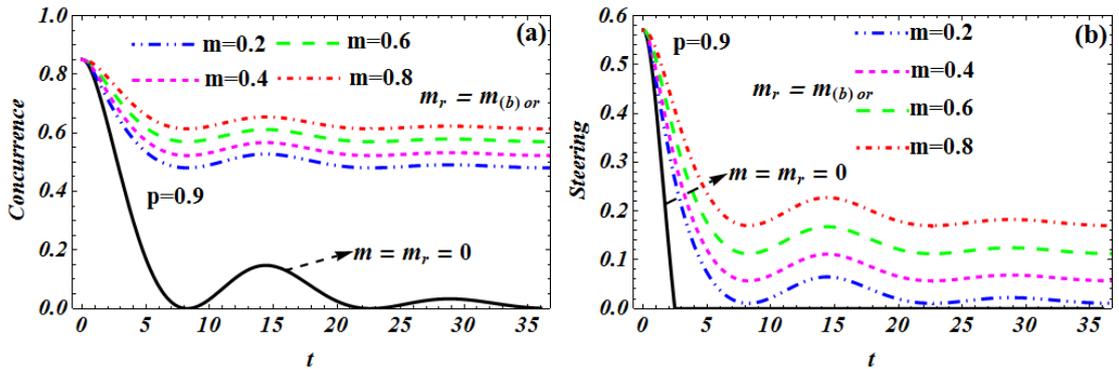

**Fig. 5** (Color online) The quantum correlations as a function of the time parameter $t$ for the different values of the weak measurement $m$ with $m_r = m_{(b)or}$ within the non-Markovian regime. Here, **(a)** shown the concurrence, and **(b)** shown the quantum steering.

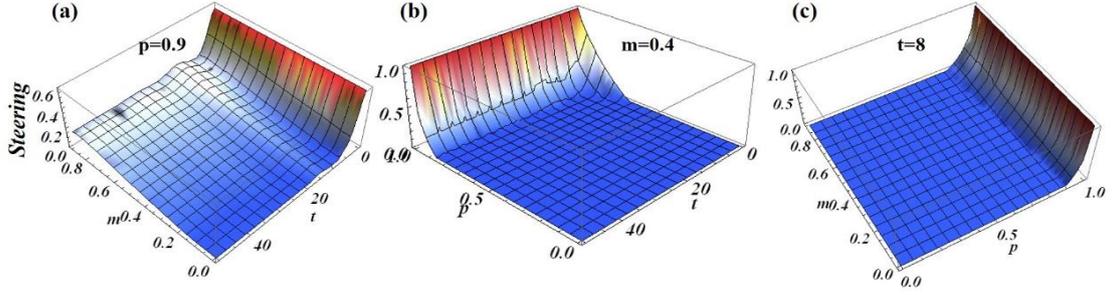

**Fig. 6** (Color online) **(a)** The quantum steering versus $m$ and $t$ for $p=0.9$ in the non-Markovian regime; **(b)** The quantum steering versus $p$ and $t$ for $m=0.4$; **(c)** The quantum steering versus $p$ and $m$ for $t=8$.

### 3.3 Bures Fidelity

Fidelity is used to develop measures of how well a quantum channel preserves information, which is involved to two different states generally [3, 64]. Fidelity of unity implies identical states whereas fidelity of zero implies orthogonal states. When some physical decoherence process occurs, the initial quantum state $\rho(0)$ will be damped to the state $\xi(t)=\rho(0)\xi$. For pure states, this measure of fidelity is generally computed using the Hilbert-Schmidt norm. However, the Hilbert-Schmidt norm is not defined for mixed states. For mixed states, a good indicator of fidelity is the Bures fidelity [65] between the damped state $\xi(t)$ and the initial state $\rho(0)$, which is defined as [65-67]

$$F = \left( tr\sqrt{\rho(0)^{1/2}\xi(t)\rho(0)^{1/2}} \right)^2. \tag{26}$$

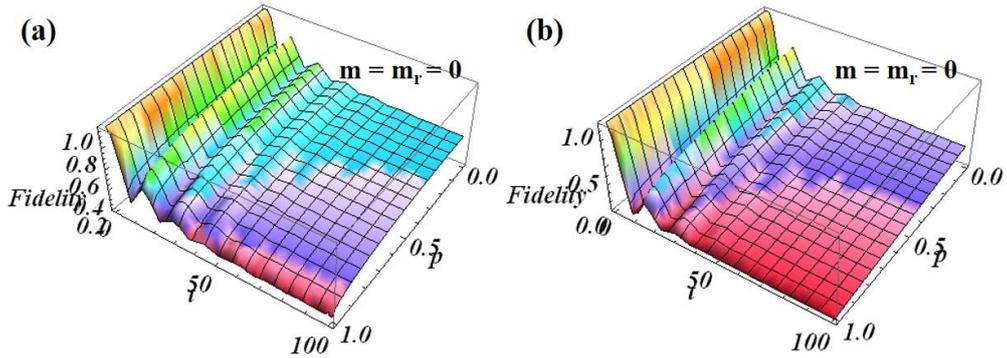

**Fig. 7** (Color online) The fidelity versus $t$ and $p$ within the non-Markovian regime. Graph **(a)** corresponds to the case 1 without WM and WMR operations, graph **(b)** corresponds to the other case without WM and WMR operations.

We first consider that the WM and WMR operations do not work. From the Fig. 7, the fidelity decreases with the increase of the time parameter *t*, and also decreases with the increase of the state parameter $p$ when the time parameter $t$ is nonzero. The results fully show that the original state is destroyed by the non-Markovian environments. Subsequently, we employ our scheme to recover the fidelity. As shown in Fig. 8, we can clearly find that the fidelity is remarkably recovered by employing quantum partially collapsing measurements. That is, our scheme can better recover the fidelity within the non-Markovian environments as well.

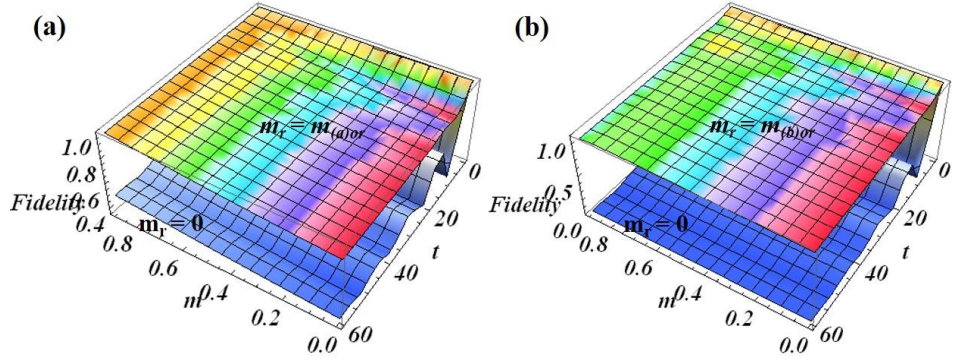

**Fig. 8** (Color online) The comparison between the fidelity of $\rho(t)_{m_r}$ and $\rho(t)_m$ as a function of $m$ and $t$ for $p = 0.9$ within the non-Markovian regime. Here, $\rho(t)_m$ is of using the WM (without WMR) operation and suffering from non-Markovian noise. Graph **(a)** corresponds to case 1, graph **(b)** corresponds to the other case.

## 4. Conclusion

To conclude, we have investigated the influence of the non-Markovian noise on quantum correlations (entanglement and steering) and how to recover the lost steerability of the quantum state. Here, two different kinds of cases (one-subsystem or all-subsystem interacts with the dissipative environments) are considered. The dynamic behaviors of entanglement and steering within the non-Markovian environment are revealed. It has been shown that both quantum entanglement and steering experience a sudden death with increasing the time parameter $t$. Besides, the dynamical interaction between quantum states and non-Markovian environment can induce the quasi-periodic entanglement revival, while quantum steering cannot regain. And the state during the resurgent entanglement cannot steer. Furthermore, we have put forward a feasible physical scheme for recovering the lost steerability of the states by making use of quantum partially collapsing measurements. The results indicate that the WM can better restore both of the degraded entanglement and steering. Moreover, the steerability of quantum states can be enhanced and the resurgent entanglement can also be employed to realize steering when the WM strength is large enough. Interestingly, not only the steerability of the quantum states can be effectively restored, but our scenario can remarkably raise the fidelity. The results also demonstrated that the larger the WM strength is, the better the effectiveness of our scheme is. Therefore, we believe that our work might be helpful to understand the dynamic behavior of quantum steering and recover the steerability of quantum states within the non-Markovian regime.

## Acknowledgments

This work was supported by the National Science Foundation of China under Grant Nos. 11575001, 61601002 and 11605028, Anhui Provincial Natural Science Foundation (Grant No. 1508085QF139), and the fund from CAS Key Laboratory of Quantum Information (Grant No. KQI201701).